\documentclass{elsart}
\usepackage{graphicx}

\def\ba{\begin{eqnarray}}
\def\ea{\end{eqnarray}}
\def\l{\label}

\def\bi{\bibitem}
\def\x{\vec{x}_\perp}
\def\p{\vec{p}_\perp}
\def\u{\vec{u}_\perp}
\def\d{\partial}
\def\nab{\vec{\nabla}}

\begin{document}

\begin{frontmatter}

\title{\bf Early evolution of transversally thermalized partons \thanksref{grant}} 
\thanks[grant]{e-mail: 
bialas@th.if.uj.edu.pl, 
\\ Mikolaj.Chojnacki@ifj.edu.pl, 
\\Wojciech.Florkowski@ifj.edu.pl \\.}

\author[uj,ifj]{Andrzej Bialas}
\author[ifj]{, \, Mikolaj Chojnacki} 
\author[ifj,as]{and Wojciech Florkowski} 

\address[uj]{M.Smoluchowski Institute of Physics, Jagellonian
University, ul. Reymonta 4, 30-059 Krak\'ow, Poland}
\address[ifj]{The H. Niewodnicza\'nski Institute of Nuclear Physics, Polish Academy of Sciences, ul. Radzikowskiego 152, 
31-342 Krak\'ow, Poland}
\address[as]{Institute of Physics, \'Swi\c{e}tokrzyska Academy,
ul.~\'Swi\c{e}tokrzyska 15, 25-406~Kielce, Poland} 

\begin{abstract} 
The idea that the parton system created in
relativistic heavy-ion collisions (i) emerges in a state with transverse momenta 
close to thermodynamic equilibrium and (ii) its evolution at
early times is dominated by the 2-dimensional (transverse) hydrodynamics of the ideal fluid
is investigated. It is argued that
this mechanism may help to solve the problem of early equilibration. 
\end{abstract}
\end{frontmatter}
\vspace{-7mm} PACS: 25.75.Dw, 25.75.-q, 21.65.+f, 14.40.-n


{\bf 1.} It is now commonly accepted that evolution of the partonic
system created in heavy-ion collisions at RHIC energies is best
described by hydrodynamics of an almost ideal fluid \cite{shplus}. In
particular, particle transverse-momentum spectra and asymmetry of the
transverse flow, as represented by the parameter $v_2$ \cite{ol}, are
reasonably well reproduced by the hydrodynamic approach.

At the same time it is also realized that the hydrodynamic picture
encounters a rather serious challenge: it requires very early thermalization of the system. 
This follows from the
fact that the asymmetry of the transverse flow is produced most
effectively at the very early stage of the evolution (when the pressure
gradients are largest). To obtain the asymmetry
parameter $v_2$ consistent with data it is necessary to start the
hydrodynamic evolution at the time below 1 fm after the collision takes
place. But application of hydrodynamics demands, as the necessary
condition, the local equilibration of the system. Such fast
equilibration is not easy to achieve with elastic perturbative
cross-sections. This puzzle of early thermalization was also widely
discussed and several exotic mechanisms were proposed for its solution
\cite{mro,strickland}. It seems fair to say, however,  that no one was yet accepted
as fully satisfactory.

In the present paper we explore the possibility that, at its early
stages, the hydrodynamic evolution applies only to transverse degrees of
freedom of the partonic system created in high-energy collisions.
The main reason of interest in such a study is the observation that the
early equilibration, if at all possible, is particularly difficult to
achieve in longitudinal direction. This is because an elastic collision
does not change significantly the direction of the colliding partons and
thus it requires very many interactions to produce a locally isotropic
distribution from the initially strongly anisotropic one \footnote{We
thank B.~Mueller for an interesting conversation about this
point.}.

At the same time one may argue that the equilibration of the partonic
spectrum in transverse direction can be obtained much easier. Indeed, it
is well known that the transverse momentum spectra observed in
nucleon-nucleon collisions are well described by the Boltzmann
distribution \cite{bec1,bec2}. As these spectra reflect - at least to some 
extent - the distribution of produced partons \cite{wn,abwc,bb}, it 
is clear that the partonic system
produced in hadronic collisions emerges already in a state close enough to
equilibrium {\it in the transverse direction} so that no much more
action is needed to achieve the goal, see e.g. \cite{bialas-string,wf-string,rafelskisteinke}. 

At this point it is important to emphasize that such a description may 
be adequate only during a certain time after the collision. Indeed, as
the time goes on, the interpartonic interactions will tend to
equilibrate the system also in longitudinal direction and thus the
standard, 3-dimensional hydrodynamics will probably take over.

To study the possibility of such purely transverse hydrodynamic
evolution and to have also a connection with physics of the relativistic
heavy-ion collisions we discuss a simple model  implementing these
ideas. The parton distribution is constructed as a superposition of many
clusters whose longitudinal motion satisfies the Bjorken in-out condition
$\eta = y$ where 
\ba
\eta=\frac12\log \frac{t+z}{t-z} 
\ea
is the spatial rapidity of the cluster and $y$ is its rapidity.
Transverse momenta in each cluster follow the equilibrium distribution
in the local rest frame of the fluid.

This is realized by the following Ansatz for the parton  distribution function 
\ba
F(x,p)=  f_\parallel  \,\, g(\tau,\eta;\x,\p), \l{ans}
\ea
where $f_\parallel$ describes the longitudinal motion of clusters,
\ba
f_\parallel = n_0 \delta(p_z t-E z) = n_0
\frac{\delta(y-\eta)}{\tau m_\perp}, \l{fpar}
\ea
with $\tau$ being the longitudinal proper time, $\tau^2=t^2-z^2$, $n_0$ is the normalization factor giving the density of clusters per unit of rapidity, and $g$ represents the distribution of transversally equilibrated partons. 
In our study we neglect parton masses and take for $g$ the Boltzmann
distribution 
\ba
g= \exp\left(-\frac{p_\mu U^\mu}{T}\right), \l{disg}
\ea
where 
$U^\mu$ is the four velocity of the fluid,
\ba
U^\mu=(u_0\cosh y,u_x,u_y,u_0\sinh y)=(u_0\cosh \eta,u_x,u_y,u_0\sinh \eta),
\ea
with $u^\mu$ being the 4-velocity of the fluid in the  frame 
where the longitudinal momentum vanishes
\ba
u^\mu=(u_0,u_x,u_y,0)=(u_0,\u,0), \;\;\; u_0^2-\u^{\,2}=1.  \l{umu}
\ea

Using these ideas, the hydrodynamic equations were derived from the 
Ansatz (\ref{ans}) -- (\ref{disg}) and the transverse expansion of the
fluid was studied numerically for various initial and final conditions. 

Starting from the initial profile determined by the density of the
wounded nucleons inside the colliding nuclei,  we have found that

(i) With the proper choice of the initial temperature $T_i$ it is possible to 
obtain parton spectra that are consistent with the experimentally observed 
pion transverse-momentum distributions.

(ii) The calculated elliptic flow parameter $v_2$ agrees with data,
provided the final temperature $T_f$ (i.e. the temperature of transition from
2-dimensional (2D) to 3-dimensional (3D) regime) is taken substantially higher
than the expected freeze-out temperature. This gives a rather short time
available for the 2D evolution.

(iii) This short evolution time implies that the transverse size of the
system is relatively small, smaller than that obtained from the HBT
measurements, thus consistently leaving space for further,
3D expansion of the system.

We thus conclude that the 2D hydrodynamics may indeed be a
reasonable description of the partonic system during
the first few fermis after the collision and it gives a good
starting point for a realistic treatment. If confirmed by
more detailed studies, this would imply serious changes in the present
understanding of the quark-gluon plasma evolution.
 
These conclusions differ substantially from those reached in a study of
a similar problem by Heinz and Wong \cite{hw}. Using a different
implementation of the ideas advocated in the present paper, they found
that it is not possible to obtain a satisfactory description of the
elliptic flow and concluded that the partonic system must be necessarily
in a 3D equilibrium from the very beginning of the evolution.
Our example shows that this is not necessarily the case. In the last
section we discuss  the possible reasons for this discrepancy.

In the next section we discuss the general form of the energy-momentum
tensor following from the Ansatz (\ref{ans}) -- (\ref{disg}) and the 
corresponding hydrodynamical equations. The numerical analysis of the 
evolution of the system are presented in Section 3. The last section 
summarizes our conclusions. 

\vspace{0.2cm}

{\bf 2.} To obtain the hydrodynamic equations we first derive, using
the Ansatz \mbox{(\ref{ans}) -- (\ref{disg}),} the general form of the
particle density and entropy density 4-vectors and of the energy-momentum 
tensor. Somewhat lenghty but straightforward calculations
give (with $\hbar=c=1$)
\ba
N^\mu=  n_0 \, \nu_g \int dy \frac{d^2p_\perp}{(2\pi)^2} p^\mu f_\parallel \, g
= n_0 \, \nu_g \frac{T^2}{2\pi\tau} U^\mu,   \l{densn}
\ea
\ba
T^{\mu\nu}=  n_0 \, \nu_g
\int dy \frac{d^2p_\perp}{(2\pi)^2} p^\mu p^\nu f_\parallel \, g
= n_0 \, \nu_g \frac{T^3}{2\pi\tau} \left[3U^\mu U^\nu -g^{\mu\nu}-V^\mu
V^\nu\right ],    \l{denst}
\ea
where $\nu_g$ is the number of internal degrees of freedom ($\nu_g$ = 16 for gluons) and 
\ba
V^\mu =(\sinh \eta,0,0, \cosh \eta)
\ea
is the 4-vector defining the longitudinal ($z$) direction.  It is also interesting to evaluate the entropy  flow,  giving
\ba
S^\mu= -n_0 \, \nu_g \int dy \frac{d^2p_\perp}{(2\pi)^2} p^\mu f_\parallel \, g
(\log g -1 ) =
n_0 \, \nu_g\frac{3T^2}{2\pi \tau} U^\mu.
 \l{denss}
\ea

From these results we also read  that the energy, entropy and particle number densities 
in the local rest frame are
\ba
 \frac{dE}{d^2x_\perp dz}=n_0 \, \nu_g\frac{T^3}{\pi \tau}\;;\;\;
\frac{dN}{d^2x_\perp dz}=
\frac13\frac{dS}{d^2x_\perp dz}=n_0 \, \nu_g\frac{T^2}{2\pi \tau} .
\ea
where the equation of state $dE/(d^2x_\perp dz) = 2P$ was used. Note that dependence of these densities on temperature is weaker than that known from 3D thermodynamics.

Using $dz=\tau d\eta$ we thus obtain the densities per unit of spatial
rapidity
\ba
\!\!\!\!\!\!\! \epsilon_2\equiv \frac{dE}{d^2x_\perp d\eta}=n_0 \, \nu_g\frac{T^3}{\pi}\;;\;\;
n_2\equiv\frac{dN}{d^2x_\perp d\eta}=n_0 \, \nu_g\frac{T^2}{2\pi} =
\frac13\frac{dS}{d^2x_\perp d\eta}\equiv \frac13 s_2, \nonumber \\
\ea
which are {\it identical} to the densities proper for 2D thermodynamics.

The hydrodynamic equations are obtained from the energy-momentum
conservation laws $\d_\mu T^{\mu\nu}=0 $ which also imply the
conservation of entropy $\d_\mu S^\mu =0 $. 
Entropy conservation implies
\ba
\d_\tau[u_0T^2]+\nab_\perp \cdot [\u T^2] =0,  \l{eqs}
\ea
while the  momentum conservation provides two other equations
\ba
\d_\tau[\u T^3 u_0]+[T^3\u] (\nab_\perp \cdot \u)
+(\u \cdot \nab_\perp)[T^3\u] +\nab_\perp T^3/3 =0. \l{eqt}
\ea
These are three equations for the three unknowns: the temperature $T$, and
two independent components of the 4-velocity $u_x$ and $u_y$.
Their characteristic feature is that they do not depend explicitly on the 
space-time variables $(\tau,x,y)$. Thus, they do not select any special 
point in space time. We have solved these equations numerically with the method described in \cite{Chojnacki:2006tv} which is a direct extension of that proposed in \cite{Baym:1983sr}. The details of the procedure are published in \cite{cf}.

\vspace{0.2cm}

{\bf 3.} To study the evolution of the system following from (\ref{eqs})
and (\ref{eqt}) we need to specify the initial conditions, adequate for
the physical situation encountered in $Au-Au$ collisions at RHIC energy.
To this end we assume that the profile (in transverse coordinates) of
the initial energy density $\epsilon_2$ is given by the density of
participants. This density is determined, for a given centrality, from
the Glauber formulae. The following discussion is given for centrality
$20-40$\%, corresponding to impact parameter $b\approx$ 7.9 fm. The
initial temperature $T_i$ at the origin, ${\vec x}_\perp = 0$, is taken as a 
free parameter.

The parton spectra are evaluated using the Cooper-Frye prescription \cite{fc}
\ba
\frac{dN}{d^2p_\perp dy} =\frac {n_0 \, \nu_g}{(2\pi)^2} \int d\Sigma_\mu(x)
p^\mu \, f_\parallel \, g, \l{cp}
\ea
where $\Sigma$ is the surface at which the 2D evolution comes to the end.
In our exercise we have taken $\Sigma$ to be the surface of constant
temperature.

\begin{figure}[t]
\begin{center}
\includegraphics[angle=0,width=0.9\textwidth]{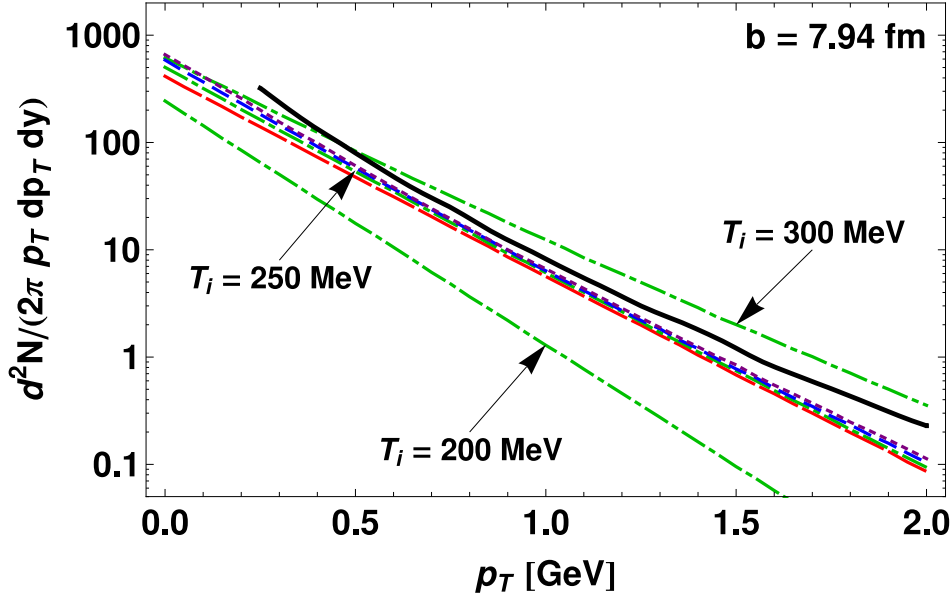}
\end{center}
\caption{\small {Transverse-momentum spectra of positive pions measured by the PHENIX Collaboration  
 in the centrality class 30-40\% (solid line) \cite{RHIC-spectra} and the model spectra of gluons for various choices of $T_i$ and $T_f$. The lowest dash dotted curve corresponds to $T_i$ = 200 MeV and $T_f$ = 180 MeV. The highest dash dotted curve was obtained for $T_i$ = 300 MeV and $T_f$ = 180 MeV. The four almost parallel lines represent our results for $T_i$ = 250 MeV and for four final values of the temperature: $T_f$ = 200 (long dashed line), 180 (dash dotted line), 160 (dashed line) and 140 (dotted line) MeV. Note that the original $\pi^+$ experimental spectra collected at $\sqrt{s_{NN}}$ = 200 GeV have been multiplied by a factor of 3 to account for the total hadron multiplicity. }}
\label{fig:spectra}
\end{figure}

The parton transverse-momentum spectrum evaluated from the model depends
substantially on the initial temperature of the system but is
practically insensitive to the final temperature (this observation was
already made in \cite{hw}). In Fig. 1, the spectrum obtained for
various values of initial temperature $T_i$ is compared with measured
spectrum of pions \cite{RHIC-spectra}. One sees that the correct slope \footnote{At this stage the experimental shape of the spectrum is not correctly reproduced since we do not include the effects such as, e.g., resonance decays and hard scattering.} is
obtained for $T_i \approx 250$ MeV. This initial temperature is used in all
subsequent calculations. The normalization, which is arbitrary in our model, was calculated with $n_0 = 1$. 

At the fixed initial temperature, results for elliptic flow are
sensitive to the value of the final temperature $T_f$. This is shown in Fig. 2
where $v_2$ calculated from the model is plotted versus $p_\perp$ for
various final temperatures. One sees that, as expected, $v_2$ grows with
decreasing $T_f$ (at a fixed $T_i$, smaller $T_f$ means longer time of
evolution and thus more time to develop the flow). As already observed
in \cite{hw}, the absence of longitudinal pressure implies that the
obtained values of $v_2$ are higher than those expected from 3D
evolution (with the same initial conditions). To bring $v_2$ close to the observed one \cite{RHIC-v2}, a fairly high $T_f \approx 180$ MeV is needed. This agrees with the point of view formulated in the first section: The final temperature should not be
interpreted as the "freeze-out" temperature but rather as the
temperature when the 2D character of the equilibrium changes into the 3D
one. This may happen well before the freeze-out and hadronization. So
the high value of $T_f$ is not surprising.

The fact the $v_2$ grows with time indicates that our approach can be adequate only at the initial stage of the evolution of the system. After this initial stage other effects such as the early transition to 3D hydro with a reduced sound velocity or with viscous effects, hadronization, and switching to transport description will reduce the growth of $v_2$ shown in Fig. 2.

\begin{figure}[t]
\begin{center}
\includegraphics[angle=0,width=0.95\textwidth]{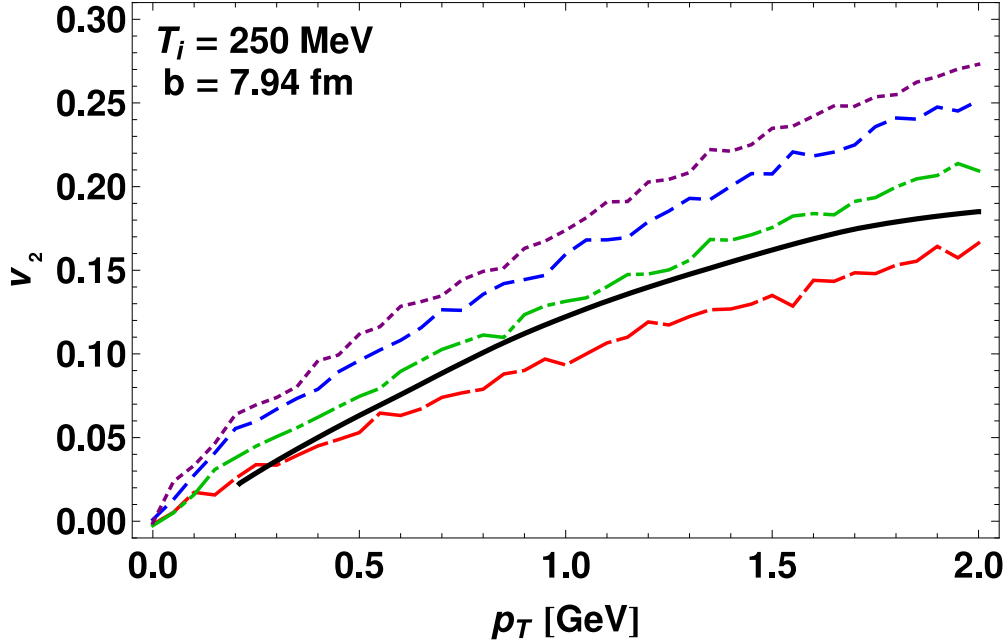}
\end{center}
\caption{\small {The elliptic flow coefficient $v_2$ as the function of the transverse momentum. The PHENIX experimental results for pions and kaons in the centrality class 20-40\% and for the collision energy $\sqrt{s_{NN}}$ = 200 GeV \cite{RHIC-v2} are compared to the model calculations with $T_i$ = 250 MeV and for four final values 
of the temperature \mbox{$T_f = 200,$} 180, 160 and 140 MeV. The curves are denoted in the same way as in Fig. 1. The best agreement is obtained for $T_f \approx$ 180 MeV.} }
\label{fig:v2}
\end{figure}

Having fixed the initial and final temperatures, we can also estimate
the life-time of the 2D evolution. It turns out to be of
about 4 fermi (for the $20-40$\% centrality class).

\vspace{0.4cm}
 
{\bf 4.} In conclusion, we have investigated the consequences of the
hypothesis that the partonic system produced in a collision of two heavy
ions is created in a state close to thermodynamical equilibrium {\it in
transverse direction}, while its longitudinal structure is
characterized by a freely-streaming collection of clusters. Two
observables were studied using the Cooper-Frye formula at the surface of
constant temperature with the initial energy density profile determined
from the density of the number of participants at a given centrality.

(a) The slope of the transverse momentum spectrum was adjusted to the
one measured for the pion spectra \cite{RHIC-spectra} giving the initial
temperature at the center of the system $T_i$ = 250 MeV.

(b) The calculated elliptic flow parameter $v_2$ gives the final temperature 
(at which the transition from 2D to 3D regime takes place) $T_f$ = 180 MeV, 
substantially higher than the expected hadronization temperature. 
The time needed for the 2D evolution was estimated to be much shorter
than that normally needed to achieve the freeze-out. This confirms the point of
view that the equilibrium changes from 2D to 3D well
before the freeze-out and hadronization.

We thus conclude that the initial period of 2D transverse
equilibration and hydrodynamic evolution of the parton system is helping
to solve the problem of early equilibration, 
 most pertinent difficulty of the present application of
hydrodynamics to physics of heavy-ion collisions. It would be certainly
interesting to investigate this possibility in more detail. 

Several comments are in order.

(i) Our conclusions differ from those obtained in \cite{hw}. As far as we can see, apart from certain technical details, there are essentially three reasons for this discrepancy. 

First, the density distributions implementing the assumption of free-streaming in the longitudinal direction differ by a factor depending on the transverse mass. This leads to rather serious consequences. Our ansatz (2) - (4) implies that at fixed rapidity the evolution of system obeys the rules of truly 2D  thermodynamics of an {\it ideal fluid}, whereas in \cite{hw} the temperature dependence of the energy and number densities follows that of 3D thermodynamics and the resulting hydrodynamic equations should be interpreted as an effective description of the {\it viscous hydrodynamics}. Our energy-momentum tensor is more symmetric and, consequently, the hydrodynamic equations differ from those used in \cite{hw}. We stress that our approach is thermodynamically consistent and the conservation laws for the energy and momentum, \mbox{$\partial_\mu T^{\mu \, \nu} = 0$}, lead to the entropy conservation, \mbox{$\partial_\mu S^\mu$ = 0}, in the way very much similar as in the standard relativistic hydrodynamics. 

Second, in Ref. \cite{hw} a new timescale parameter $\tau_0$ is introduced. This parameter -- absent in our formulation -- plays an important role in the argument used in \cite{hw} to reject a two-dimensional evolution. The large values of $\tau_0$, found in the fitting procedure by Heinz and Wong, are in their opinion inconsistent with the very idea of transverse thermalization because such thermalization may take place only at the beginning of the evolution of the system, when the considered times $\tau$ are small (with $\tau_0 < \tau$). The replacement $\tau_0 \to n_0/m_\perp$ (which formally transforms their  model into our approach) solves this problem if $n_0$ is interpreted as the density of 2D clusters in rapidity.

Third, the authors of \cite{hw} confront their results for the elliptic flow with the earlier 3D hydrodynamic calculations, whereas we fix our parameters by comparison with the present data. Our results shown in Fig. 2 indicate that we can reproduce the experimental values of $v_2$ already at high values of the temperature where most of the effect is created. The 3D hydro results shown in Fig. 7 of \cite{hw} are above the data, hence they cannot be used as the reference point for rejecting the concept of transverse thermalization.

(ii) In  our calculations we have assumed a sharp transition between the
2D and a possible 3D evolution. This simplification,
necessary to obtain a reasonably tractable problem, can be removed in
future, more sophisticated analyses. This raises interesting questions
about the nature of this transition and about the final fate of the
clusters.

(iii) It should be clear that our investigation represents only the first
step towards a fully realistic description. Many details, as e.g. the mechanism of 2D $\to$ 3D transition, the relative distribution of clusters, their internal (longitudinal) structure, the rapidity dependence of the system, are left for future
work. Nevertheless, we feel that we produced some compelling arguments for the
existence of a substantial period of purely transverse thermodynamic
equilibrium and hydrodynamic evolution of the partonic system (in the form of the ideal fluid) at the very beginning of the collision process.

\vspace{0.3cm}
{\bf Acknowledgements}

Discussions with W. Czy\.z, B. Mueller, St. Mr\'owczy\'nski, and K. Zalewski are highly appreciated. This investigation was partly supported by the MEiN research grant
1 P03B 045 29 (2005-2008) and Polish Ministry Of Science and Higher Education grants Nos. N202 153 32/4247 and N202 034 32/0918.

\end{document}